\documentclass[10pt,pra,twocolumn,showpacs,amsfonts,letterpaper]{revtex4}
\usepackage{graphicx}
\usepackage{bm}
\usepackage{times}
\newcommand{\xosc}{x_{\rm{osc}}}
\newcommand{\kosc}{k_{\rm{osc}}}

\newcommand{\modname}{Tonks }
\newcommand{\modabr}{Tonks }

\newcommand{\xp}{{\tilde x}}

\newcommand{\pphi}{{\tilde \phi}}
\newcommand{\ppsi}{{\tilde \psi}}

\newcommand{\xv}{x}

\newcommand{\q}{x}

\newcommand{\threea}{\left\{\int_{-\infty}^{\xv}
   -\int_{\xv}^{\xv'} + \int_{\xv'}^{\infty}
  \right\}}

\newcommand{\threeb}{\left\{\int_{-\infty}^{\xv'}
   -\int_{\xv'}^{\xv} + \int_{\xv}^{\infty}
  \right\}}

\newcommand{\aln}[1]{\alpha_{#1}(\xv,\xv')}

\newcommand{\sgn}{\text{sgn}(\xv'-\xv)}
\newcommand{\erf}{\text{erf}}

\newcommand{\eref}[1]{(\ref{#1})}
\begin{document}
\title {Momentum distribution for a one-dimensional trapped gas of hard-core bosons}
\author{G.J. Lapeyre, Jr.}
\email{lapeyre@physics.arizona.edu}
\author{M.D. Girardeau}
\email{girardeau@optics.arizona.edu}
\author{E.M. Wright}
\email{Ewan.Wright@optics.arizona.edu}
\affiliation{Optical
Sciences Center and Department of Physics, University of Arizona,
Tucson, AZ 85721}
\date{\today}
\begin{abstract}
Using the exact $N$-particle ground state wave function for a
one-dimensional gas of hard-core bosons in a harmonic trap we
develop an algorithm to compute the reduced single-particle
density matrix and corresponding momentum distribution. Accurate
numerical results are presented for up to $N=8$ particles, and the
momentum distributions are compared to a recent analytic
approximation.
\end{abstract}
\maketitle
\section{\label{sec:intro}Introduction}
Recent advances in atom waveguide technology
\cite{vengalattore01,cren02,leanhardt02,Schreck2001a,key00,thywissen99,mueller99,dekker00,bongs01,hinds98}
and the realization of Bose-Einstein condensates in optical
and magnetic traps of variable aspect ratio
\cite{Greiner2001a,Goerlitz01} have spurred interest in the
properties of degenerate quantum gases in lower dimensions. In
particular, the \modname gas, in which strong
transverse confinement and low temperature and density allow the
gas to be modeled as a one-dimensional (1D) system of point
particles with hard-core interactions \cite{olshanii98,petrov00},
is of considerable theoretical interest due to the fact that it
defies a mean-field description, but is on the other hand exactly
soluble via the Fermi-Bose mapping \cite{girardeau60,girardeau65}.
Although the exact many-body wave function can be written in a compact
form using the mapping theorem, obtaining information about important
observables has proven to be a difficult task. One such quantity that
bears the signature of the \modabr gas is the momentum distribution,
which has a sharp peak at zero momentum \cite{olshanii98}, in
contrast to the Fermi sea for the corresponding 1D gas of fermions. In
a mathematical {\it tour de force}, Lenard \cite{lenard64} obtained
upper bounds on the momentum distribution for a homogeneous \modabr
gas, and some elaborations of that work followed \cite{vaidya79}. In a
previous paper \cite{girardeau01a} we obtained numerical results for
the momentum distribution of a harmonically trapped \modabr gas for up
to $N=10$ particles, and Minguzzi {\it et al.} \cite{minguzzi02}
developed an analytic approximation for the high-momentum tail of the
momentum distribution of a trapped gas.  Cazalilla \cite{Cazalilla02}
 obtained an analytic approximation for the momentum distribution
of \modabr gas confined in a box using a description of the system as
a Luttinger liquid.

Our previous calculations of a trapped \modabr gas were performed
using a Monte Carlo (MC) integration of the many-body wave function to
obtain the single-particle reduced density matrix from which the
momentum distribution was obtained via Fourier
transformation. Although the data thus generated were useful, an
improved method is desirable because of the limited accuracy of MC
integration.  Even these MC data were limited to $N=10$ particles,
which required weeks of computer time.  It seems clear that the way
forward is to have high-precision numerical data available for testing
the validity of approximations. In this paper we start from the
$N$-particle ground state wave function for a one-dimensional
condensate of hard-core bosons in a harmonic trap and develop an
algorithm to compute the reduced single-particle density matrix and
momentum distribution.  The key advantage of this approach is
that, although we are limited to only $N=8$ particles at present by
computer resources, these data are of high precision, thus providing a
testing ground for analytic approximations.

In Sec. \ref{sec:ground}, we give a precise definition of the
system and find the ground state wave function using the Fermi-Bose
mapping theorem. In Sec. \ref{sec:reduced}, we write the
single-particle reduced density matrix $\rho_1$ and develop a method
for its numerical solution.  In Sec. \ref{sec:mom} we use these
results for $\rho_1$ to evaluate the momentum distribution and compare
the results to a recent approximation for the high-momentum tail.
\section{\label{sec:ground}Ground state wave function}
The Hamiltonian of $N$ bosons in a 1D harmonic trap is
\begin{equation}\label{eq1}
\hat{H}=\sum_{j=1}^{N}
\left[-\frac{\hbar{^2}}{2m}\frac{\partial^2}{\partial \xp_{j}^{2}}
+\frac{1}{2}m\omega^{2}\xp_{j}^{2}\right]  .
\end{equation}
We assume that the two-body interaction potential consists only of
a hard-core of 1D diameter $a$. This is conveniently treated as a
constraint on allowed wave functions
$\ppsi(\xp_{1},\ldots,\xp_{N})$ such that
\begin{equation}\label{eq2}
\ppsi=0\quad\text{if}\quad |\xp_{j}-\xp_{k}|<a, \quad 1\le j<k\le N,
\end{equation}
rather than as an infinite interaction potential. It follows from
the Fermi-Bose mapping theorem
\cite{girardeau60,girardeau65,girardeau00} that the exact $N$-boson
ground state $\ppsi_{B0}$ of the Hamiltonian (1) with the
constraint (2) is
\begin{equation}\label{eq3}
\ppsi_{B0}(\xp_{1},\ldots,\xp_{N})=|\ppsi_{F0}(\xp_{1},\ldots,\xp_{N})|,
\end{equation}
where $\ppsi_{F0}$ is the ground state of a fictitious system of $N$
spinless fermions with the same Hamiltonian (\ref{eq1}) and
constraint.  At low densities it is sufficient
\cite{olshanii98,petrov00} to consider the case of impenetrable point
particles, the zero-range limit $a\to0$ of Eq. (2).  Since wave
functions of ``spinless fermions'' are antisymmetric under coordinate
exchanges, their wave functions vanish automatically whenever any
$\xp_{j}=\xp_{k}$, the constraint has no effect, and the corresponding
fermionic ground state is the ground state of the {\em ideal} gas of
fermions, a Slater determinant of the lowest $N$ single-particle
eigenfunctions $\pphi_n$ of the harmonic oscillator (HO)
\begin{equation}\label{eq4}
\ppsi_{F0}(\xp_{1},\ldots,\xp_{N})=\frac{1}{\sqrt{N!}}
\det_{(n,j)=(0,1)}^{(N-1,N)}\pphi_{n}(\xp_{j}) .
\end{equation}
The HO orbitals are
\begin{equation}\label{orbital}
\pphi_{n}(\xp)= \frac{1}
{\pi^{1/4}\xosc^{1/2}\sqrt{2^{n}n!}}e^{-\q^{2}/2}H_{n}(\xp/\xosc),
\end{equation}
with $H_n(x)$ the Hermite polynomials and $\xosc=\sqrt{\hbar/m\omega}$
the ground state width of the harmonic trap for a single atom.  For
convenience, we introduce the dimensionless coordinates
$\q_i=\xp_i/\xosc$, and the corresponding ground state wave function
$\psi_{B0}$.  As we have shown in previous work \cite{girardeau01a},
substitution of Eq. (\ref{orbital}) into Eq. (\ref{eq4}) and some matrix
manipulations \cite{aitken51} lead to a simple but exact expression of
the Bijl-Jastrow pair product form for the $N$-boson ground state:
\begin{equation}\label{groundstate}
\psi_{B0}(x_{1},\ldots,x_{N})=C_{N}\left[\prod_{i=1}^{N}e^{-\q_{i}^{2}/2}
\right]
\prod_{1\le j<k\le N}|x_{k}-x_{j}|,
\end{equation}
%
with normalization constant
\begin{equation}
C_{N}=2^{N(N-1)/4}
\left[N!\prod_{n=0}^{N-1}n!\sqrt{\pi}\right]^{-1/2}.
\end{equation}
\section{\label{sec:reduced}Single-particle density matrix}
\subsection{Analytic formula}
The reduced single-particle density matrix with normalization
$\int\rho_{1}(x,x)dx=N$ for the ground state given by
Eq. (\ref{groundstate}) is
%
\begin{eqnarray}
  \rho_{1}(x,x')&=&N\int\psi_{B0}(x,x_{2},\ldots,x_{N})\nonumber \\
 && \times \psi_{B0}(x',x_{2},\ldots,x_{N})dx_{2}\cdots dx_{N}
\nonumber\\
&  = & {\mathcal N}_{N}e^{-\q^{2}/2}e^{-{\q'}^{2}/2} I(\q,\q'),
\label{rho1}
\end{eqnarray}
where the integration is from $-\infty$ to $\infty$ for each
coordinate unless otherwise stated, and
%
\begin{equation}
{\mathcal N}_N = N 2^{N(N-1)/2} \pi^{-N/2}
   \left[ \prod_{n=0}^N n! \right]^{-1},
\end{equation}
and we have defined
\begin{eqnarray}
 I(\q,\q') &=& \int\prod_{i=2}^{N}e^{-\q_{i}^{2}}
       |\q_{i}-\q||\q_{i}-\q'| \nonumber \\
 && \times \prod_{2\le j<k\le N}
   (\q_{k}-\q_{j})^{2}d\q_{2}\cdots d\q_{N}.
\label{rhoint}
\end{eqnarray}
In the following subsection, we develop a method for analyzing
$I(\q,\q')$. We will see that, when $N$ is small enough (say $2$ or
$3$), the exact expression is manageable, but that we must turn to
numerical methods for larger $N$.
\begin{figure*}[!]
\includegraphics*[width=15cm, angle=0]{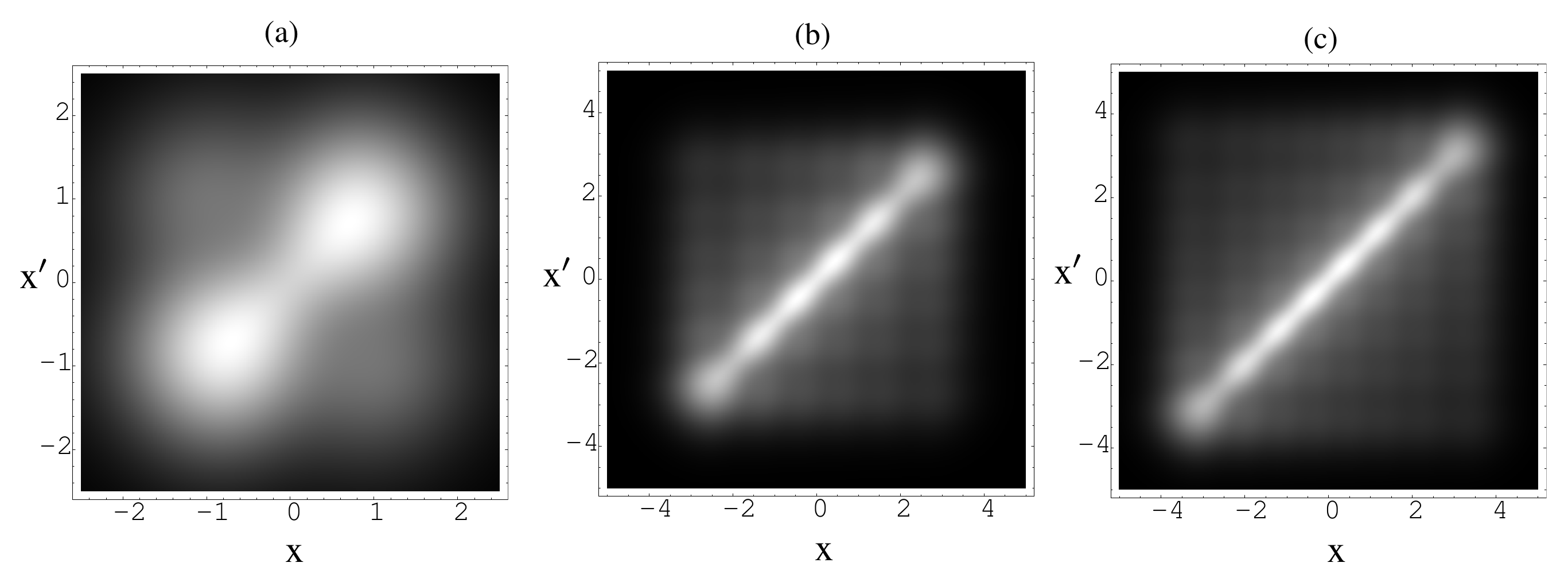}
\caption{Gray scale plots of the dimensionless reduced
density matrix $\xosc\rho_1(\q,\q')$ as a function of the
dimensionless
coordinates $\q$ and $\q'$, for (a) $N=2$, (b) $N=6$, and (c) $N=8$. }
\label{Fig:one}
\end{figure*}
\subsection{Numerical approach}
The multidimensional integral (\ref{rhoint}) can be expressed in
terms of polynomials, Gaussians, and error functions. But, even for
relatively small $N$, the number of terms in such an expression is too
large to be useful when written. We previously evaluated the reduced
single-particle density matrix using
MC methods \cite{girardeau01a}, but with limited numerical accuracy. Here we
present a seminumerical approach in which we represent the integral
in terms of incomplete gamma functions and evaluate the result to
machine precision for particular values of $\xv$ and $\xv'$.

We reduce the integral $I(\xv,\xv')$ to incomplete gamma functions in
the following way.  Consider the case $\xv<\xv'$. We first integrate
over $\xv_2$, writing
\begin{eqnarray}
 I(\xv,\xv') &=& \int d\xv_{3}\cdots d\xv_{N} \,
  I_2 \prod_{i=3}^{N}e^{-\xv_{i}^{2}}|\xv_{i}-\xv||\xv_{i}-\xv'| \nonumber \\
  &&\times  \prod_{3\le j<k\le N}(\xv_{k}-\xv_{j})^{2} ,
\end{eqnarray}
where
%
\begin{eqnarray}\label{twoint}
 I_2 &=& I_2(\xv,\xv',\xv_3,\ldots,\xv_N) \nonumber \\
  &=& \threea P_2 \, d\xv_2,
   \quad \xv'<\xv
\end{eqnarray}
and
\begin{eqnarray}\label{Ptwo}
 P_2&=&P_2(\xv,\xv',\xv_2,\ldots,\xv_N) \nonumber \\
  &=& e^{-\xv_2^2}(\xv_2-\xv)(\xv_2-\xv')\prod_{2<k\le N} (\xv_2-\xv_k)^2.
\end{eqnarray}
The integrand $P_2$ is an analytic function of $\xv,\xv'$, and $\xv_2$
( in the sense that derivatives of all orders in these variables
exist), so that the integral over each of the three intervals is
analytic in these variables.  Furthermore, we can evaluate the
integral over $\xv_2$ in Eq. (\ref{twoint}) easily because the integrand
is a Gaussian multiplied by a polynomial.  We next integrate over
$\xv_3$, getting
\begin{eqnarray}
 I(\xv,\xv') &=& \int d\xv_{4}\cdots d\xv_{N} I_3
  \prod_{i=4}^{N}e^{-\xv_{i}^{2}}|\xv_{i}-\xv||\xv_{i}-\xv'| \nonumber \\
 && \times \prod_{4\le j<k\le N}(\xv_{k}-\xv_{j})^{2},
\end{eqnarray}
where
\begin{eqnarray}\label{threeint}
 I_3 &=& I_3(\xv,\xv',\xv_4,\ldots,\xv_N) \nonumber \\
  & =& \threea I_2 P_3 \, d\xv_3,
       \quad \xv'<\xv
\end{eqnarray}
and
\begin{eqnarray}
 P_3 & = & P_3(\xv,\xv',\xv_3,\ldots,\xv_N)\nonumber \\
  &= & e^{-\xv_3^2}(\xv_3-\xv)(\xv_3-\xv')\prod_{3<k\le N} (\xv_3-\xv_k)^2.
\end{eqnarray}

It is important to note that, because $I_2$ is a polynomial in
$\xv_3,\ldots,\xv_N$, the integral $I_3$ in Eq. (\ref{threeint}) can be
solved by the same technique used to solve $I_2$ in Eq. (\ref{twoint}).
We continue this procedure, defining $I_4$, etc., until all of
the integrals are finished.  At each stage, one has a more complicated
polynomial in the remaining independent variables in the integrand.
Before continuing, we note that there is, of course, nothing
essentially new when we choose $\xv'<\xv$. For instance, for $I_2$, we
have
\begin{equation}
 I_2 =  \threeb P_2 \, d\xv_2, \quad \xv'<\xv.
\end{equation}
If we examine each of the integrals $I_2, I_3, \ldots$ above, we see
that all of the integrals to be computed can be reduced to terms
proportional to integrals of the form
\begin{equation}\label{alone}
\aln{n} = \threea e^{-u^2} u^n \, du, \quad \xv<\xv'
\end{equation}
and
\begin{equation}\label{altwo}
\aln{n} = \threeb e^{-u^2} u^n \, du \quad \xv>\xv'.
\end{equation}

We now present an algorithm for expressing the integrals $I_2,\dots$
in terms of the $\aln{n}$ defined in Eqs. (\ref{alone}) and
(\ref{altwo}). To achieve this for $I_2$, we take $P_2$, drop the
factor of $\exp(-\xv_2^2)$, expand the remaining factors as a
polynomial in $\xv_2$, and replace each occurrence of $\xv_2^n$ with
$\aln{n}$. The result is $I_2$ expressed as a polynomial in
$\xv,\xv',\xv_3,\ldots$ with coefficients involving the $\aln{n}$.
This expression for $I_2$ is substituted into Eq. (\ref{threeint}) and the
same procedure is then used to compute $I_3$, and so on, until all
powers of $\xv_m^n$ have been replaced by $\aln{n}$. This procedure
can be simplified by performing all of these substitutions at once.
To this end, consider
\begin{equation}\label{fullpoly}
 P =  \prod_{i=2}^{N}
   (\xv_i-\xv)(\xv_i-\xv')
  \prod_{2\le j<k\le N} (\xv_j-\xv_k)^2 .
\end{equation}
To compute the integral $I(\xv,\xv')$ given in Eq. (\ref{rhoint}), we
expand Eq. (\ref{fullpoly}) and substitute $\aln{n}$ for each occurrence
of $\xv_m^n$, for any $m$.  In addition, for each value of $m$ for
which a term is independent of $\xv_m$, a factor of $\aln{0}$ must be
included. The result is $I(\xv,\xv')$ expressed as a polynomial in the
$\aln{n}$ for approximately $2N$ values of $n$.  Rather than print the
results, we store a table of the coefficients of the powers of
$\aln{n}$ on a computer. Then, a table of the $\aln{n}$ is computed
for a particular pair $\xv,\xv'$, and $I(\xv,\xv')$ is computed using
this table together with the table of coefficients.
Evaluating the $\aln{n}$ using numerical integration is relatively
inefficient. Instead, we evaluate them numerically using well-known,
efficient routines to compute incomplete gamma functions.
We define an indefinite integral
\begin{equation}\label{Fdef}
 F_n(\xv) = \int e^{-\xv^2} \xv^n.
\end{equation}
Then using definitions \eref{alone} and \eref{altwo} we
have
\begin{eqnarray}\label{alphadef}
  \aln{n} &=& F_n(\infty)-F_n(-\infty)\nonumber \\
 &  + &\text{sgn}(\xv'-\xv)2[F_n(\xv)-F_n(\xv')].
\end{eqnarray}
$\aln{n}$ is continuous, but has a cusp at $\xv=\xv'$.
For numerical computation, we use
\begin{eqnarray}
 &F_n(\infty)&-F_n(-\infty) = \int_{-\infty}^{\infty} e^{-u^2} u^n \nonumber \\
  &=& \cases{ 
     0  & \text{ if $n$ is odd} ,  \cr \noalign{\vskip 7pt}
     {\displaystyle \Gamma\left(\frac{1+n}{2}\right)=\sqrt\pi
       \frac{(n-1)!!}{2^{n/2}} } & \text{otherwise}, \cr }
\end{eqnarray}
and
\begin{eqnarray}
 2[F_n(\xv)-F_n(\xv')] &=&
   [\text{sgn}(\xv)]^n \Gamma\left(\frac{1+n}{2},\xv^2\right)\nonumber \\
    & - &[\text{sgn}(\xv')]^n \Gamma\left(\frac{1+n}{2},\xv'^2\right).
\end{eqnarray}

\subsection{Examples}

In this section we carry out in detail the algorithm given above for
two particles and give the result for three particles.  We first
choose $N=2$ and tabulate the required values of $F_n$ and $\aln{n}$.
Integrating Eq. (\ref{Fdef}) by parts, we write
\begin{eqnarray}
 F_0(\xv)&=&\frac{\sqrt\pi}{2}\erf \xv,  \nonumber \\
 F_1(\xv) &=& -\frac{1}{2}e^{-\xv^2}, \nonumber \\
 F_2(\xv) &=& \frac{1}{4}\sqrt\pi \text{erf}\xv -\frac{1}{2}\xv e^{-\xv^2}. \nonumber
\end{eqnarray}
Using Eq. \eref{alphadef} we then have
\begin{eqnarray}
\aln{0} &=& \sqrt\pi + \sgn \sqrt\pi \left( \erf \xv - \erf \xv'\right), \nonumber  \\
\aln{1} &=& -\sgn\left(e^{-\xv^2}- e^{-{\xv'}^2} \right), \nonumber  \\
\aln{2} &=&  \frac{\sqrt\pi}{2} + \sgn
    \bigg[\frac{\sqrt\pi}{2}\left( \erf \xv - \erf \xv'\right) \nonumber \\
       &&+\xv'e^{-\xv^2} - \xv e^{-{\xv'}^2} \bigg]. \nonumber
\end{eqnarray}
Applying the algorithm outlined in the previous section we find for two particles
\begin{widetext}
\begin{eqnarray}\label{algtwo}
 \rho_1(\xv,\xv') &=& {\mathcal N}_2 e^{-\frac{\xv^2}{2}-\frac{{\xv'}^2}{2}}
   \int_{-\infty}^{\infty}e^{-\xv_2^2}
   \left|(\xv_2-\xv)(\xv_2-\xv') \right| d\xv_2 \nonumber 
    =  {\mathcal N}_2 e^{-\frac{\xv^2}{2}-\frac{{\xv'}^2}{2}}
     \big[\aln{2}-(\xv+\xv')\aln{1} \nonumber 
     +\xv\xv'\aln{0}\big] \nonumber \\
   &=&  {\mathcal N}_2 e^{-\frac{\xv^2}{2}-\frac{{\xv'}^2}{2}}
    \Bigg\{\sqrt\pi\left[\frac{1}{2}+\xv\xv' \right] \nonumber 
    +\sgn \left[\left(\frac{1}{2}+\xv\xv'\right) \sqrt\pi
     \bigg( \erf \xv - \erf \xv'\right) \nonumber 
    + \xv'e^{-\xv^2} - \xv e^{-{\xv'}^2}  \bigg]
         \Bigg\}.
\end{eqnarray}
\end{widetext}

For $N=3$, one can easily compute the expression for $I(\xv,\xv')$ in terms
of $\aln{n}$ by hand, with the result
\renewcommand{\a}{\alpha}
\begin{eqnarray}
\rho_1(\xv,\xv') &=&  2{\mathcal N}_3 e^{-\frac{\xv^2}{2}-\frac{{\xv'}^2}{2}}
   \big\{-\a_3^2 +\a_2\a_4 \nonumber \\*
  &+& p^2(-\a_2^2 +\a_1\a_3) + t(\a_2^2 - 2\a_1\a_3 +\a_0\a_4)\nonumber \\*
  &+&t^2(-\a_1^2 +\a_0\a_2) \nonumber \\*
  &+&p\big[ \a_2\a_3 -\a_1\a_4
    + t(\a_1\a_2-\a_0\a_3) \big] \big\},\nonumber
\end{eqnarray}
where $p=\xv+\xv'$, $t=\xv\xv'$, and the explicit dependence of $\a_n$
on $\xv$ and $\xv'$ is omitted for clarity.  For larger values of $N$,
$I(\xv,\xv')$ rapidly becomes more difficult to compute by hand. In
the next section we present the results of carrying out the algorithm
detailed above on a computer.
\subsection{Numerical results}
We have evaluated the above integrals numerically for $N=2$--$8$.

Figure \ref{Fig:one} shows a gray scale plot of the dimensionless
reduced single-particle density matrix $\xosc\rho_1(\q,\q')$
versus the normalized coordinates $\q$ and $\q'$ for (a) $N=2$, (b)
$N=6$, and (c) $N=8$. We verified that along the diagonal
$\rho_1(\q,\q)=\rho(\q)$ reproduced the single-particle density
\cite{kolomeisky00}.
\section{\label{sec:mom}Momentum distribution}
%
%
In terms of the boson annihilation and creation operators in position
representation (quantized Bose field operators) the one-particle
reduced density matrix is
\begin{equation}
\rho_{1}(x,x')=\langle\Psi_{B0}|\hat{\psi}^{\dagger}(x')\hat{\psi}(x)|
\Psi_{B0}\rangle.
\end{equation}
The momentum distribution function $n(k)$, normalized to
$\int_{-\infty}^{\infty} n(k)dk=N$, is
$n(k)=\langle\Psi_{B0}|\hat{a}^{\dagger}(k)\hat{a}(k)|\Psi_{B0}\rangle$
where $\hat{a}(k)$ is the annihilation operator for a boson with momentum
$\hbar k$. Then
\begin{equation}\label{momint}
n(k)=(2\pi)^{-1}\int_{-\infty}^{\infty}dx\int_{-\infty}^{\infty}dx'
\rho_{1}(x,x')e^{-ik(x-x')}.
\end{equation}
The spectral representation of the density matrix then leads to
$n(k)=\sum_{j}\lambda_{j}|\mu_{j}(k)|^2$ where the $\mu_j$ are Fourier
transforms of the natural orbitals:
$\mu_{j}(k)=(2\pi)^{-1/2}\int_{-\infty}^{\infty}\phi_{n}(x)e^{-ikx}dx$.
\begin{figure}
\includegraphics*[width=\columnwidth,angle=0]{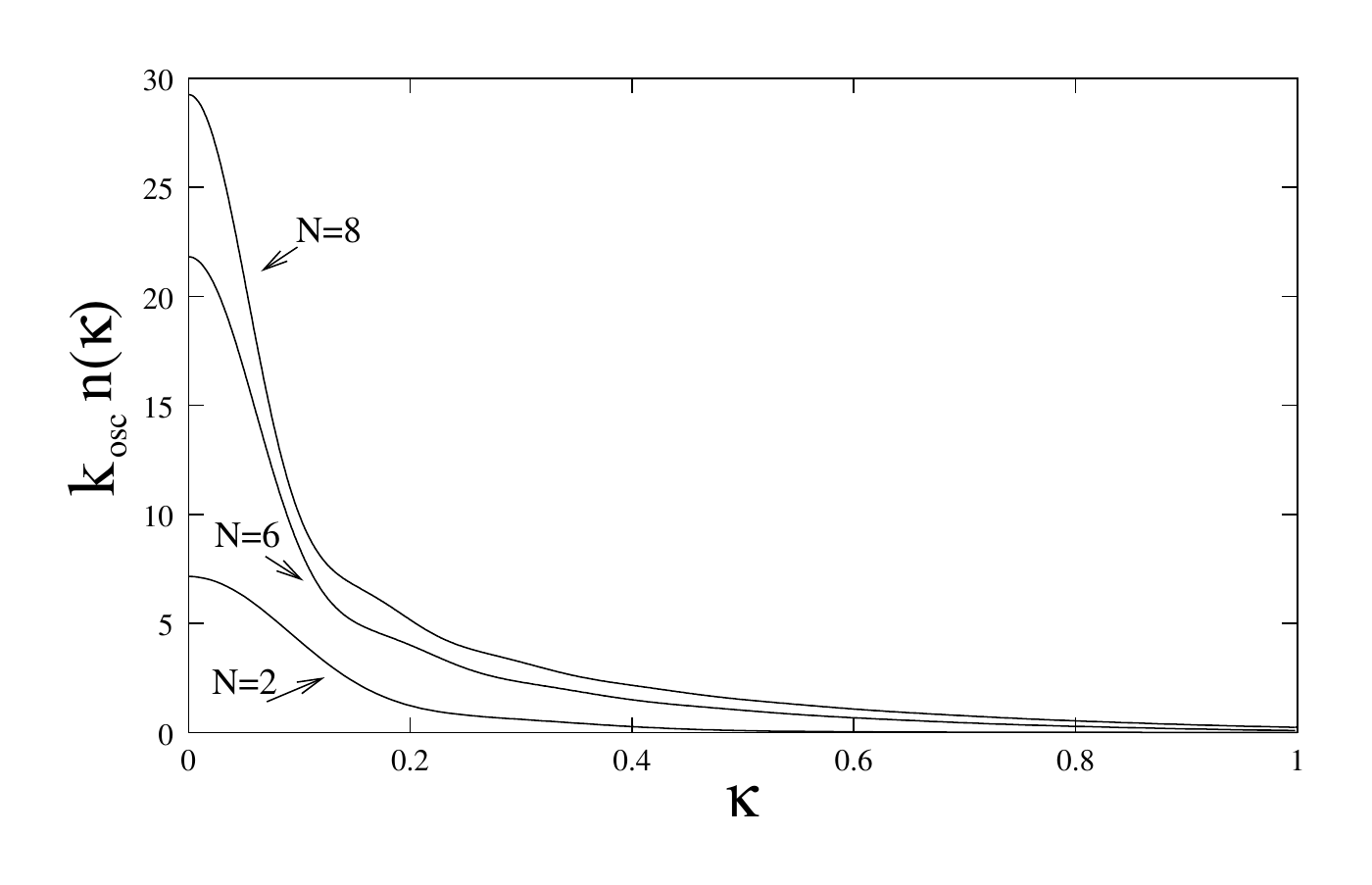}
\caption{Dimensionless momentum distribution $\kosc n(\kappa)$ versus
normalized momentum $\kappa=k/\kosc$ for  $N=2$, $N=6$, and
$N=8$.  Note the peaks becoming sharper with increasing $N$.}
\label{Fig:two}
\end{figure}
\begin{figure}
\includegraphics*[width=\columnwidth,angle=0]{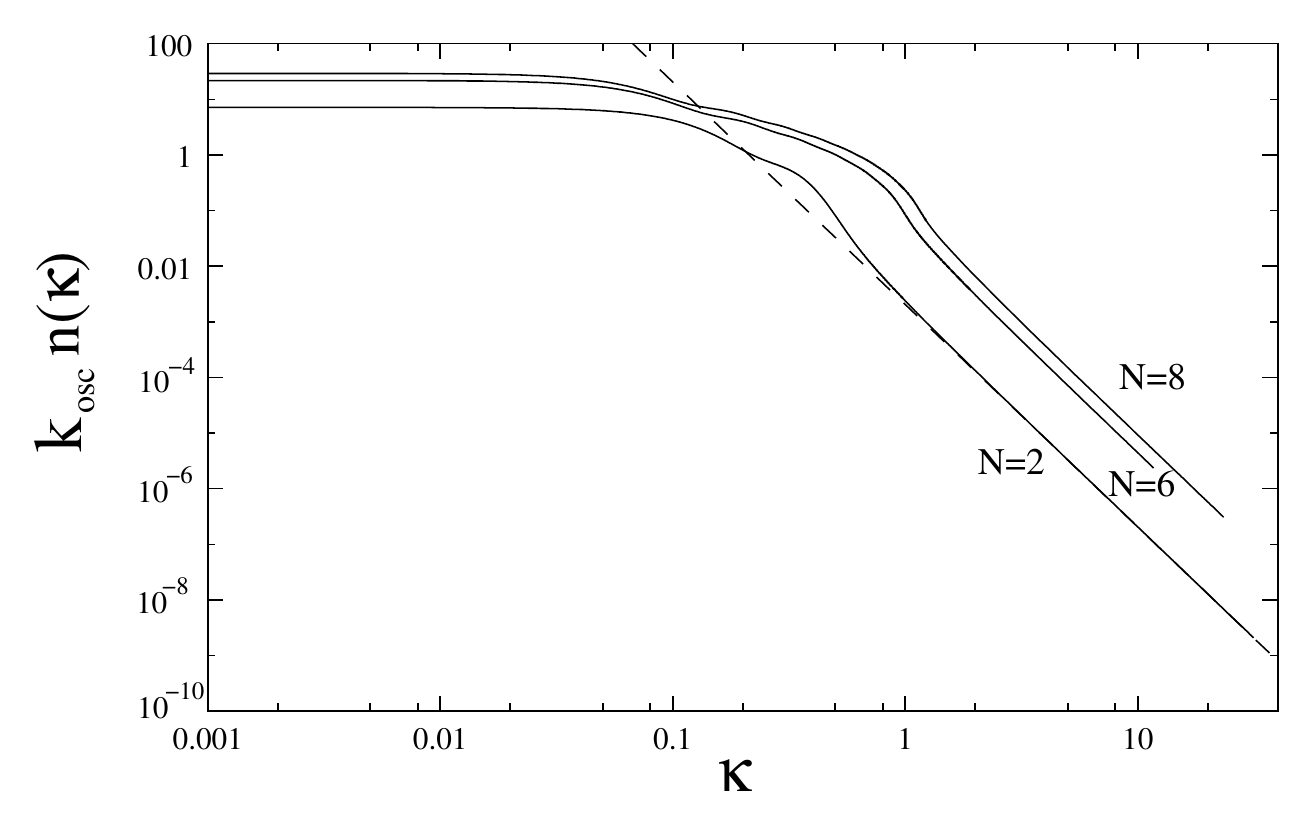}
\caption{Dimensionless momentum distribution $\kosc n(\kappa)$
versus normalized momentum $\kappa=k/\kosc$ on log-log scale.
 The dashed line is the asymptotic expression given
 by Eqs. (\ref{asympdecay}) and (\ref{twoasym}). }
\label{Fig:three}
\end{figure}
Figure \ref{Fig:two} shows the numerically calculated dimensionless
momentum distribution $\kosc n(\kappa)$ versus normalized momentum
$\kappa=k/\kosc$, with $\kosc=2\pi/\xosc$, for (a) $N=2$, (b) $N=6$, and
(c) $N=8$. We typically evaluated $\rho_1(\q,\q')$ to machine precision
for smaller values of $\kappa$, with precision decreasing to a part in
$10^{-6}$ for the largest values of $\kappa$.  The key features are
that the momentum distribution maintains the peaked structure reminiscent
of the spatially uniform case \cite{olshanii98,lenard66} for the 1D
HO, and that the peak becomes sharper with increasing atom number $N$.
This is to be expected since as the number of atoms increases the
many-body repulsion causes the system to become more spatially uniform
within the trap interior.
Minguzzi \textit{et al.} \cite{minguzzi02} determined that the momentum
distribution for finite $N$ decays according to
\begin{equation}\label{asympdecay}
 \lim_{k\to\infty} k^4 n(k) = A_N,
\end{equation}
where $A_N$ depends only on the number of particles.
In particular, for $N=2$, they found
\begin{equation}\label{twoasym}
 A_2 = 2 \sqrt{\frac{2}{\pi}} (\hbar m \omega)^{-3/2}.
\end{equation}
Figure \ref{Fig:three} shows our numerical results for $N=2$
approaching this asymptotic form.
The dashed line in Fig. \ref{Fig:three} shows
$k_{osc}^5A_2/\kappa^4$ versus $\kappa=k/k_{osc}$, and we see that
this approximation agrees with our numerical results for $N=2$ for
high momenta. Furthermore, inspection of the numerical results for
other values of $N$ shows a $1/k^4$ dependence in the high-momentum tail.

\section{\label{sec:summary}Summary and conclusions}
In summary, we have developed a method for obtaining high-accuracy
results for the momentum distributions of trapped \modabr gases, and
presented results for up to eight particles. Our results agree
reasonably with the high-momentum approximation $n(p)\propto 1/p^4$
obtained by Minguzzi {\it et al.}
%
\begin{acknowledgments}
This work was supported by Office of Naval Research
Grant No. N00014-99-1-0806 and the U.S. Army Research Office.
\end{acknowledgments}
\bibliography{bose1d,BEC}
\end{document}